\documentclass{article}

\usepackage{url}
\usepackage{graphicx}
\usepackage{amsmath}
\usepackage{amssymb}

\usepackage{mathtools}
\usepackage{mathrsfs}
\usepackage{enumerate}
\usepackage{amsthm}
\usepackage{algorithm}
\usepackage{algpseudocode}
\usepackage{setspace}
\usepackage{authblk}
\usepackage{float}

\usepackage[round]{natbib}

\newcommand{\naive}{na\"{\i}ve}
\newcommand{\Naive}{Na\"{\i}ve}

\DeclareMathOperator*{\argmax}{arg\,max}

\newcommand{\E}[1]{\mathrm{E}[#1]}

\renewcommand{\v}[1]{\mathrm{Var}[#1]}

\newcommand{\set}[1]{\left\{ #1 \right\}}
\renewcommand{\epsilon}{\varepsilon}

\newcommand{\s}{s^*}
\newcommand{\x}{x^*}
\newcommand{\n}{{n^*}}
\renewcommand{\r}{r^*}

\newcommand{\D}{{D^*}}

\newcommand{\reals}{\mathbb{R}}
\newcommand{\m}{\mathcal{M}}
\newcommand{\K}{\mathcal{K}}

\newcommand{\I}{\mathcal{I}}

\begin{document}

\title{\textbf{Optimal-Design Domain-Adaptation for Exposure Prediction in Two-Stage Epidemiological Studies}}

\author{
  Ron Sarafian \\
  Department of Industrial Engineering, Ben Gurion University of the Negev, Be'er Sheva, Israel\\
  \texttt{ronsarafian@gmail.com}
  \and
  Itai Kloog\\
  Department of Geography and Environmental Development, Ben Gurion University of the Negev, Be'er Sheva, Israel\\
  \and
  Jonathan D. Rosenblatt \\
  Department of Industrial Engineering, Ben Gurion University of the Negev, Be'er Sheva, Israel
}
\date{}

\maketitle


\begin{abstract}
{
In the first stage of a \emph{two-stage study}, the researcher uses a statistical model to impute the unobserved exposures.
In the second stage, imputed exposures serve as covariates in epidemiological models.
Imputation error in the first stage operate as measurement errors in the second stage, and thus bias exposure effect estimates. 

This study aims to improve the estimation of exposure effects by sharing information between the first and second stage.

At the heart of our estimator is the observation that not all second-stage observations are equally important to impute. 
We thus borrow ideas from the optimal-experimental-design theory, to identify individuals of higher importance.
We then improve the imputation of these individuals using ideas from the machine-learning literature of domain-adaptation.

Our simulations confirm that the exposure effect estimates are more accurate than the current best practice.
An empirical demonstration yields smaller estimates of PM effect on hyperglycemia risk, with tighter confidence bands.

Sharing information between environmental scientist and epidemiologist improves health effect estimates.
Our estimator is a principled approach for harnessing this information exchange, and may be applied to any two stage study. 
}
\end{abstract}


\section{Introduction} \label{sec:introduction}
Environmental epidemiology (EEPI) is a branch of epidemiology that studies the health effects of environmental exposures, such as air pollution and temperature.
Studies are typically observational, and relate exposure measurements to health outcomes at the individual-level.
Common EEPI studies are usually conducted at the individual-level, and exposure assessment is geolocated to a particular point, such as an individual's residence \citep{montero2015spatial, hodges2013richly}.

Defining and measuring exposure accurately is far from trivial.
Inaccurate exposure assessments will introduce measurement error, and will bias \emph{exposure effect} (EE) estimates \citep{szpiro2011efficient}. 
At the individual level, \emph{direct} measurement of ambient exposure is usually impossible, and thus, an \emph{indirect} measurement is performed. 
This indirect measurement can be thought of as an \emph{imputation}, or \emph{prediction} of the unknown  exposure. 
A common approach to indirect measurement is predicting exposure from satellite imagery.
For instance, indirect measurements of \emph{fine particulate matter} (PM\textsubscript{2.5}) are obtained by fitting a model that takes satellite products like \emph{aerosol optical depth} (AOD) as predictors.
The model is first calibrated using fixed ground monitoring stations, then used to predict PM\textsubscript{2.5} where ground stations are unavailable but AOD is \citep{shtein2018estimating, sarafian2019gaussian}.

In a \emph{two-stage study} the EE is estimated using indirect measurement, where the locations of individuals differ from the locations of monitors.
Two-stage studies are tremendously popular in the EEPI community \citep{szpiro2013measurement}. 
The current paradigm in EEPI research is that exposures are predicted once, and then serve multiple studies.
Namely, in the first stage, the statistician provides a single exposure surface (e.g., a grid);
then different epidemiologists use it for different studies.

The above imputation introduces measurement error.
The measurement-error literature offers many remedies \citep{carroll2006measurement}.
For instance, \citet{gretton2009covariate, spiegelman2010approaches, szpiro2011efficient, lopiano2011comparison}, compare various error correction methods.
For satellite predictions, \citet{just2018correcting} use machine-learning algorithms to improve predictions and thus reduce measurement error.
The above either treat predicted exposures as fixed, or improve them uniformly for all studies. 
This, in contrast to the current proposal, where we tailor predictions to each study. 

Ignoring the subsequent epidemiological study is the current practice in exposure modeling.
Yet, Diao, and 16 other respectable epidemiologists, statisticians, and exposure scientists suggest that it should not be the case \citep{diao2019methods}.
In the context of PM\textsubscript{2.5}, they recommend that researchers from both communities work together when applying exposure predictions into health impact assessments.
A similar argument was voiced by \citet{szpiro2011does}, which recommend that the development of models for exposure prediction and EE estimation should be considered simultaneously.

In this work, we try to answer these calls and suggest a framework for improving exposure predictions for EEPI studies.
For improvement, we focus on the accuracy of EE estimates. 
This is quite different from today's state-of-the-art exposure models, which focus on first-stage (exposure) predictions.

The following example illustrates our argument:
Assume the epidemiologist is fitting a linear model. 
We know that in simple linear regression, $ Var[\hat \beta]=\sigma^2/\sum (x_i-\bar x)^2 $, where $ \hat \beta $ is the estimated slope, $ x_i $ the observed predictors, and $ \sigma^2 $ the variance of departures from the linear trend. 
We thus see that for accurate effect estimates, extreme values of $ x_i $ are more influential than mean values.
This suggests that the environmental epidemiologist should pay more attention to accurate predictions of $ x $ at its extremes than its typical values. 

Unlike a designed experiment, the epidemiologist is not free to choose the exposure levels of the participants.
However, in two-stage studies, the epidemiologist is free to improve the accuracy for ``important'' individuals. 
Identifying individuals of importance can be done with the theory of \emph{optimal-design}.
Having identified these individuals, we need to improve their exposure predictions. 
This can be done with the theory of \emph{domain-adaptation}, a.k.a.\ \emph{transduction} or \emph{dataset-shift}.
Our approach thus consists of two components:
(i) At the second-stage: use optimal-design to identify the observations of importance.
(ii) At the first stage, use domain-adaptation to improve exposure predictions where required for the second stage.
The optimal-design might depend on the exposures themselves, so an iterative structure is suggested.

Here, we cast our ideas as an estimation algorithm.
We present simulation results that confirm its efficiency comparing to current (non-iterative) best practices, and apply it to empiricaly estimate PM\textsubscript{2.5} effect on hyperglycemia risk.

\section{Methods} \label{sec:methods}
\subsection{Problem Setup} \label{sec:psetup}

Let $y \in \mathcal{Y}$ be a health outcome. 
Usually $Y \subset \reals$, or $Y= \{0,1\}$. 
We allow the exposure to be multivariate (e.g., PM\textsubscript{2.5} and PM\textsubscript{10}; temperature and humidity; etc.), and denote it with $x \in \mathcal{X} \subset \reals^p$.
Lastly, $z \in Z \subset \reals^q$ are some other covariates in the epidemiological model.
Typically, we assume that $y$'s distribution, $ F $, depends on $ x $, and $ z $, via some parameter $\theta$:
\begin{align}\label{eq:model}
    y|x,z \sim F(\theta (x,z) ).
\end{align}
We will denote all the modeling assumptions in Eq.(\ref{eq:model}) with $ \m $.
For instance, $ \m $ may state 
$ y|x,z \sim \mathcal{N} (x' \beta_x + z' \beta_z, \sigma^2 )$. 
Here $ F $ is Gaussian, and $\theta = (\beta_z, \beta_x, \sigma^2)'$. 
The EE in this case is $\beta_x$.

The epidemiologist has access to a sample of $i = 1,...,n$ individuals.
For each $i$, $y_i$ and $z_i$ are known, but the true exposure, $ x_i $ is unknown.
We denote with $ \hat x_i $ an indirect measurement of the exposure of individual $i$, sometimes known as an \emph{imputation}, \emph{surrogate}, or \emph{proxy}.
We call $D$ the \emph{second-stage dataset}, that consists of $n$ triplets $(y_i,\hat x_i,z_i)$.
Because $ \hat x_i $ is used instead of $ x_i $, the epidemiologist may estimate  $ \hat \beta_{\hat x} $, but not $ \hat \beta_{x}$. A good proxy of $x$ is one where $\Vert \hat \beta_{\hat x} - \beta_x \Vert \approx 0$.

Moving to the exposure scientist.
We use $ (.)^* $ to denote measurements in the first stage. 
We think of $ \x $ as the exposure at location $ \s $, which is measured directly without noise; for instance, because $ \s $ is the location of a ground monitoring station.
We think of $ \r $  as a vector of geographical covariates that includes satellite products (e.g. AOD), and other spatial predictors such as land-use, spline-basis functions, etc.
Because $ \x,\r $ are spatial we may use $ x(\s),r(\s)$ when we want to emphasize the spatial structure of the problem.
The data $ \D $, consists of $ \n $ pairs $(\x_i,\r_i)$. 
The indirect measurement, $ \hat x(s) $, is done by applying some statistical learning algorithm, on the \emph{training dataset} $ \D $, and applying the resulting predictor $h_{\D}$ at $ s $.
We denote this $ \hat x(s):=h_\D(r(s)) $.

To fix ideas, $ h_\D $ may be a Gaussian random field \citep{sarafian2019gaussian}, a Gradient boosting machine \citep{just2020advancing}, a deep network \citep{park2020estimating}, or an ensemble of different machine-learning algorithms \citep{hough2020multi}.

In this text we assume the first-stage learning paradigm is \emph{empirical risk minimization} (ERM).
Let $h$ be a predictor from some \emph{hypothesis class},
and $l(.)$ the loss for predicting $h(r)$ instead of $x$: 
$ l \big( h(r), x \big) \mapsto \reals^+ $.
The risk function can be written as:
\begin{align} \label{eq:risk}
    \mathcal{R}(h) := \mathbb{E} \big[l\big(h(r(\s)), x(\s)\big)\big].
\end{align}
A more explicit formulation that clarifies the sources of variability in Eq.~\ref{eq:risk} is given in Section~\ref{sec:adaptation}.

In a usual two-stage study, the exposure scientist's problem is finding the predictor that minimize empirical risk on the training set:
\begin{align} \label{eq:independent}
    \hat h_\D^{0} := \arg\min_{h} \ \frac{1}{\n} \sum_{i = 1}^{\n} l(h(r(\s_i)), x(\s_i)).
\end{align}
We denote the risk-minimizing predictor with $h^{0}$, where $0$ is the number of iterations between the exposure scientist and the epidemiologist.

Alternatively, the exposure scientist may adopt the epidemiologist's view, and seek a predictor that has a small second-stage estimation error:
\begin{align} \label{eq:oracle}
    h^{orc} := \arg\min_{h}  \mathbb{E} \Big[ \Big( \hat \beta_{\hat x} \big( h(r(\s_i) \big)  - \beta_x \Big)^2 \Big].
\end{align}
We call $h^{orc}$ an \emph{oracle}, because it requires knowledge of the unknown $\beta_x$. 

Ideally, $\hat h_\D^{0}$ and $h^{orc}$ are close, and return similar predictions.
In reality, $\hat h_\D^{0}$ and $h^{orc}$ might be quite different, which will inflate $ \Vert \hat \beta_{\hat x}-\beta_x \Vert  $ at the second stage.

We will now focus on \emph{generalized linear model} (GLM), for two reasons.
First, they are very popular second-stage models in EEPI studies.
Second, optimal-design for GLMs have been extensively studied \citep{dean2015handbook}.
Extensions to the GLM are discussed in Section~\ref{sec:departing-from-glm}.

A GLM has the following components:
(1) An assumed distribution of health outcome's, $F$, that is restricted to the exponential family. We denote its mean by $ \mu $.
(2) A linear predictor of exposure and other covariates: 
$\eta = \beta' \Phi(x,z) $, where $\beta \subset \theta$ are unknown parameters.
(3) A \emph{link function} $g(\mu) = \eta$ relating the linear predictor to the expected health outcome.

The vector $\Phi(x,z)$ is the transformed features. 
Usually, $\Phi(x,z) = (1,x,z)'$, i.e., a $1+p+q$ vector, but $\Phi(x,z)$ may also include interactions and other transformations of covariates.
In a GLM, the variance of $y$ has the form 
\begin{align} \label{eq:var_y}
    \v{y|x,z} = \phi \, V(\mu),
\end{align}
where $\phi$ is a constant and $V(.)$ is some function of $\mu$, both determined by $F$.

\subsection{Optimal Design} \label{sec:odesign}
For ease of exposition, we assume that the second-stage model includes only the ($p$-dimensional) exposure as explanatory variables with an intercept, so the linear predictor is given by 
$\eta = \beta' \Phi(x) = \beta_0 + x' \beta_x$.
In Section~\ref{sec:second-stage-covariates} we discuss the scenario where more covariates are included.
An optimal design identifies the values of $x$, and their sampling probabilities, so that uncertainty in estimates is minimal.
These are called the \emph{support points} of the design, and the corresponding probabilities are called the \emph{design weights}.
For $ j=1,\dots,J $ denote $x_j\in \mathcal{X}$ an exposure support point, and $w_j$ a corresponding weight.
The design weights meet: $\sum_{j=1}^J w_j = 1$.
A design for a second stage is a collection of $J$ support points and design weights: 
$ \xi:=\set{(x_j,w_j)}_{j=1}^J $.

One of the nice properties of the GLM with design $ \xi $ is the general and compact form of the information matrix, $\I(\xi, \beta)$, it implies:
\begin{align} \label{eq:I}
    \I(\xi, \beta) = \sum_{j=1}^J w_j \, u(x_j) \, \Phi(x_j)\Phi'(x_j),
\end{align}
where $u(x_j)$ is called the \emph{model weight} of point $ x_j $:
\begin{align} \label{eq:u}
    u(x_j) := \frac{1}{\phi \, V(\mu_j)} \Big( \frac{\partial \mu_j}{\partial \eta_j}\Big)^2,
\end{align}
and $V(.)$ is the variance function defined in Eq.~\ref{eq:var_y}  \citep{dean2015handbook}. 

An optimal design, $ \tilde \xi $, minimizes some function $ \Psi $ of the information, with respect to support point and weights, within the permissible region:
\begin{align} \label{eq:opt}
    \tilde \xi := \arg\max_{\xi}  \Psi \big( \I(\xi;\beta) \big).
\end{align}

Setting $\Psi(.) := \det(.)$ is known as a \emph{D-optimal} design, which is arguably the most popular;
$\Psi(.) := \text{Tr}(.)$ is an \emph{A-optimal} design; 
$\Psi(.) := \lambda_{min}(.)$, where $\lambda_{min}(.)$ returns the minimum eigenvalue is an \emph{E-optimal} design.
A review is given in \citet[][Chp. 2]{fedorov2013optimal}.

In linear models, $\tilde \xi$ is independent of any unknowns.
In non-linear models, $\tilde \xi$ depend on both the unknown $\beta$ and the assumed distribution.
Clearly, an inconvenient circularity arises if the optimal design depends on the quantity to be recovered by the design.
The literature offers various remedies to this matter \citep{dean2015handbook}.
For instance, in \emph{sequential designs} the idea is to start from an initial static design to estimate the parameters of interest;
then sequentially parameters and design are updated.

\subsection{Adaptation to the Optimal Design} \label{sec:adaptation}
\emph{Domain-Adaptation} deals with the learning of a predictor in one population, and its application in another. 
These are known as \emph{source} and \emph{target} distributions, respectively. 
In our case, we will learn to predict exposure in the first stage, and apply predictions in the second stage. 
This means, for instance, that the first (second) stage marginal distribution of $ x $ represents exposures at monitoring stations (residences of subjects).
We denote the joint source distribution of geographical covariates and exposure with $P_S(r,x)$.
We use $P_T(r,x)$ for their joint target distribution.
Our goal is to train a predictor with samples from $P_S(r,x)$, to predict the unknown exposure of samples drawn from $P_T(r,x)$.
$P_T(r,x)$ and $P_S(r,x)$ might be quite different.
For instance, when individuals are exposed to different levels than those measured in monitoring stations.

We can now write Eq.~\ref{eq:risk} more precisely, i.e., the risk function that the exposure scientist minimizes if unaware of the epidemiologist's needs:
\begin{align} \label{eq:soureRisk}
    \mathcal{R}_S(h) &:= \int l\big( h(r), x \big) \, d P_S(r,x).
\end{align}

In $\mathcal{R}_S(h)$, integration is taken w.r.t the source distribution.
If, on the other hand, the exposure scientist is aware of the target domain, then instead of Eq.~\ref{eq:soureRisk}, the risk is the average loss in the target population:
\begin{align} \label{eq:targetRisk}
    \mathcal{R}_T(h) &:= \int l\big( h(r), x \big) \, d P_T(r,x).
\end{align}

$\mathcal{R}_T(h)$ can be rewritten using integration w.r.t the source distribution:
\begin{align} \label{eq:riskIW}
    \mathcal{R}_T(h) &= \int \omega(r,x) \, l\big(h(r), x \big) \, d P_S(r,x),
\end{align}
where $\omega(r,x)$ are called the \emph{importance weights} \citep{shimodaira2000improving}, and satisfy: 
\begin{align} \label{eq:omega}
    \omega(r,x) = \frac{P_T(r,x)}{P_S(r,x)}.
\end{align}

When $P_T(r,x) = P_S(r,x)$, i.e. $\omega(r,x) = 1$, then no adaptation is required.
This is actually implicit in today's current best practices. 
In contrast, when $P_T(r,x) \neq P_S(r,x)$, the exposure scientist should know $\omega(r,x)$ in order to minimize the empirical counterpart of $\mathcal{R}_T(h)$.

The issue is simplified if we assume a \emph{prior-shift} (a.k.a \emph{label shift}) \citep{quionero2009dataset}.
Under the prior-shift assumption, conditional distributions are assumed equal: 
\begin{equation}\label{eq:prior-shift}
	P_S(r|x) = P_T(r|x),
\end{equation}
whereas the prior distributions of the exposures differ: $P_S(x) \neq P_T(x)$.
More on this assumption in Section~\ref{sec:prior-shift}.
Substituting the prior shift assumption in Eq.~\ref{eq:omega} implies:
\begin{align} \label{eq:iw}
    \omega(r,x) = \omega(x) = \frac{P_T(x)}{P_S(x)}.   
\end{align}

$P_S(x)$ is the marginal distribution of exposures in the first-stage.
We estimate it from $ D $ and denote the estimator $\hat P_S(x)$.
$ P_T(x) $ is the marginal distribution of exposures in the target population, i.e., in the second stage. 
In \citet{sarafian2020domain} $ P_T(.) $ was known because we set it to be the locations of residence of second-stage subjects.
In this contribution, we set the target to be the exposure support points identified by the second-stage's OD.

To set $ P_T(x) $, we observe that the optimal design, $\tilde \xi$ is a (discrete) probability distribution over $ \mathcal{X} $.
We could define $ P_T(x)=\tilde{\xi}(x) $ and $ \omega(x)=\tilde \xi(x) /P_S(x)$.
However, $\tilde \xi$ and $P_{T}(x)$ do not agree on the support, thus, we suggest smoothing $ \tilde \xi(x) $, for instance, with kernel density estimators.

By convolving $ \tilde{\xi} $ with some kernel function $ \K $, we get a mixture distribution:
\begin{align} \label{eq:PT}
    \hat P_{T}(x;\tilde \xi) \propto \tilde \xi(x) \circledast  \K(x,x_j) =  \sum_{j=1}^J w_j \, \K(x,x_j),
\end{align}
where $\K(x,x_j)$ is some kernel function that weights according to the distance of $x$ from the support point $x_j$, and $\propto$ means equality up to some normalizing constant.

The empirical counterpart of the weighted risk in Eq.~\ref{eq:riskIW} can now be derived, using 
$\hat \omega (\x) := \hat P_{T}(\x;\tilde \xi) / \hat P_S(\x)$.
We denote with $h_\D^{odiw}$ the \emph{optimal-design importance-weighted} (ODIW) predictor, which minimize this empirical risk:
\begin{align} \label{eq:odiw}
    h_\D^{odiw} = \arg\min_{h} \ \frac{1}{\n} \sum_{i = 1}^{\n} \hat \omega(x(\s_i)) \, l(h(r(\s_i)), x(\s_i)).
\end{align}

\subsection{Algorithm} \label{sec:algo}
Equipped with optimal-design and domain-adaptation theory, we now suggest an estimation algorithm that allows to estimate the EE accurately.
The crux is to iterate between the exposure scientist (\underline{Es}ther) and epidemiologist (\underline{Ep}hraim):
(1) Esther provides exposure predictions. 
(2) Ephraim uses them to estimate the EE.
(3) Ephraim uses optimal-design to mark data points of importance.
(4) Esther uses domain-adaptation to improve predictions at those points.

The details of our algorithm are the following.
Denote $h_{\D,\omega}$ the ODIW predictor from Eq.~\ref{eq:odiw}, learned with importance weights $\omega$ and dataset $ \D $.
Denote with $\hat x \gets h_{\D,\omega}$ the $n$-vector of exposures predicted using $h_{\D,\omega}$.
Denote with $\omega^0 $ an initialization of weights.
Denote $\hat \beta_{\hat x} \gets \m_{\hat x} $, effect estimates in epidemiological model $ \m $, estimated using exposures $ \hat x $.
Denote with 
$ \tilde \xi \gets \argmax_\xi \{ \det ( \I(\xi; \hat \beta_{\hat x}) ) \}$
the D-optimal design of $\m$, as defined in Eq.~\ref{eq:opt}, with information matrix $\I$ evaluated at $\hat \beta_{\hat x}$, as defined in Eq.~\ref{eq:I}.
Finally, let $\hat P_{T}(x;\tilde \xi)$ be the estimated density implied by $\tilde \xi$ as defined in Eq.~\ref{eq:PT}, and $\hat{P}_S(x)$ a density estimate for $x$ in $\D$.

\begin{algorithm}[h] \caption{ODIWI Estimator} \label{algo:iterator}
	\begin{algorithmic}
	\Function{EE Estimator}{$ \D, D, \m, \K, \omega^0$}
	\State $\hat{x}^1 \gets h_{\D,\omega^0}$ 
	\Comment{Initialize exposures}
	
	\For{$l \in \set{1,...,L}$}
	\State $\hat \beta_{\hat x^l} \gets \m_{\hat x^l}$ 
	\Comment{Estimate EE with current exposures}
	
	\State 
	$\tilde \xi^l \gets \argmax_\xi \{ \det ( \I(\xi; \hat \beta_{\hat x^l}) ) \}$
	\Comment{Find D-optimal design}
	
	\State $w^l_i \gets \hat P_{T}(\x_i; \tilde{\xi}^l) / \hat{P}_S(\x_i), \forall i = 1,...,\n$
	\Comment{Re-Weight $ x_i $}
	
	\State $\hat{x}^{l+1}\gets h_{\D,\omega^l}$
	\Comment{Update exposures using current weights}
	
	\EndFor 
	
	\State \Return $\hat \beta^L$ 
	\EndFunction
	\end{algorithmic}
\end{algorithm}

Algorithm~\ref{algo:iterator} has many design choices.
The obvious ones are the epidemiological model, $ \m $, and the predictor's hypothesis class.
These have received enough attention in the literature so we will not elaborate. 
Design choices that are more specific to our setup include:
initialization choices;
optimization tuning;
stopping rules;
and the optimality criterion.
These are discussed in Section~\ref{sec:disc_algo}.

\section{Results}
\subsection{Simulation Analysis} \label{sec:sim}
The following simulation examines the accuracy of Algorithm~\ref{algo:iterator} in estimating the EE.
For first-stage data, $\D$, we simulate satellite data and spatial predictors, from a zero-mean multivariate normal distribution: $r_i \sim \mathcal{N}(0,\Sigma)$, where $\Sigma = U'U$ and $U$'s entries are independent uniformly distributed: $U_{i,j} \sim Unif[0,1]$.
The true exposure, $x$, is some linear function of $r$ with additive Gaussian noise: $x_i = \gamma' r_i + \epsilon_i$.
A binary health outcome, $y_i$, is simulated from a Bernoulli distribution with a logit link:
\begin{align} \label{eq:logistic}
    P(y_i=1|x_i; \beta) = \frac{\exp (\beta_0 + \beta_x x_i)}{1 + \exp (\beta_0 + \beta_x x_i)}.
\end{align}

Predicted exposures are restricted to be linear in $ r $.
We also compared other classes, including support vector regression with non-linear kernels; 
results were qualitatively the same, and so not reported herein.

\begin{figure}[H] 
\centering
  \includegraphics[width=1\textwidth]{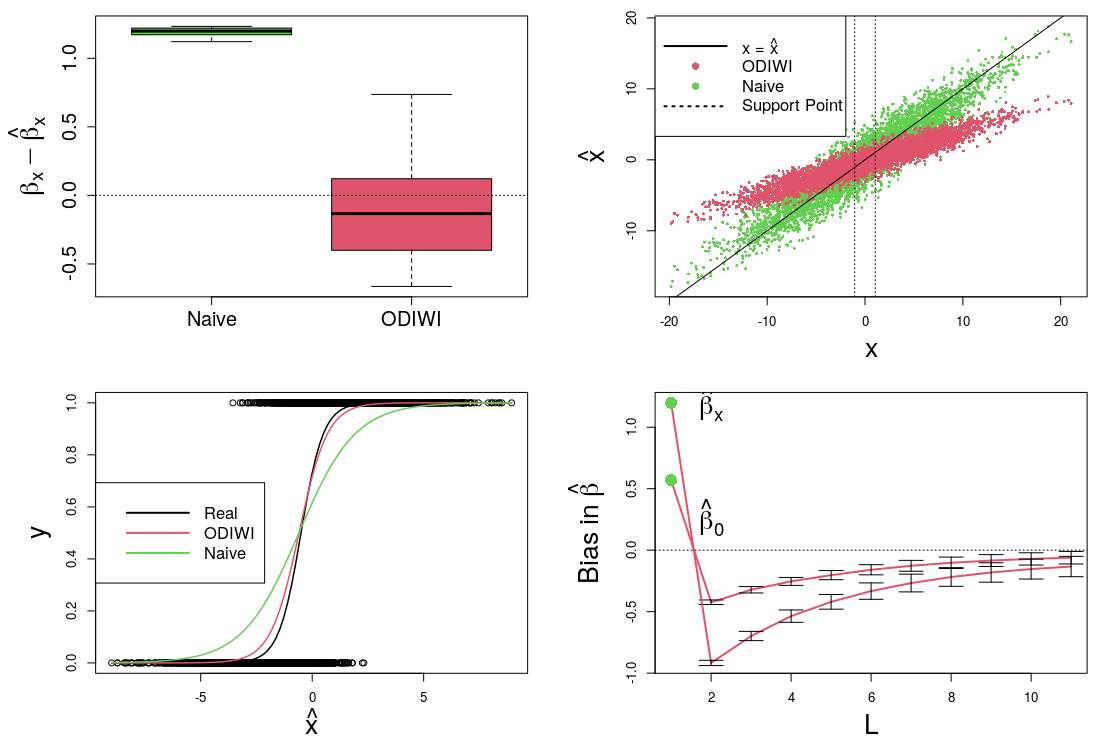}
  \caption{\footnotesize 
  	Simulation results. 
  \textbf{Top-left:} $ \hat \beta_{\hat x}-\beta_x $ of \Naive \ and ODIWI estimators. 
  \textbf{Top-right:} $x$ against $\hat{x}$ of \Naive \ (green) and ODIWI (red) in a single realization, after $L=10$ iterations.
  Optimal design ($ \tilde{\xi} $) in vertical dashed lines. 
  \textbf{Bottom-left:} real probabilities (black), \Naive \ estimates (green), ODIWI estimates (red), in a single realization.
  \textbf{Bottom right:} $ \E{\hat \beta_{\hat x}-\beta_x} $ along iterations.
}
  \label{fig:sim1}
\end{figure}

Figure~\ref{fig:sim1} compares the accuracy of Algorithm~\ref{algo:iterator}'s ODIW-Iterative estimator (ODIWI) and the standard two-stage estimator (\Naive) in estimating $\beta_x$.
Our main finding is that a small number of iterations almost always improves accuracy compared to a non iterative approach. 
This can be seen from the distribution of $ \beta_x-\hat \beta_{\hat x} $ in the upper left display.
The number of iterations seems to decrease $ \beta_x-\hat \beta_{\hat x} $ (lower right), but after enough iterations, overfitting may kick in, and errors will grow (not reported).

The upper right panel illustrates the true exposure ($ x $) versus predicted ($ \hat x $) after $L=10$ iterations.
It is not surprising that the \Naive \ predictions are fairly accurate for all $ x $ (green).
ODIWI, on the other hand, does not try to give accurate predictions for all $ x $, but rather, only at the support of the optimal design ($ \tilde \xi $, in red).
Interestingly, ODIWI gives worse $ \hat x $ (on average), but improves 
$ \hat \beta_{\hat x}-\beta_x $.

\begin{figure}[H] 
\centering
  \includegraphics[width=0.85\textwidth]{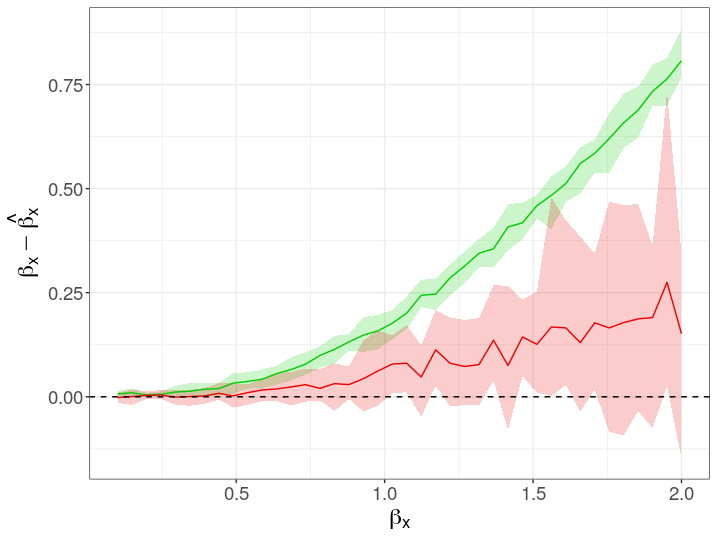}
  \caption{\footnotesize
   $\hat \beta_{\hat x}-\beta_x $ for the \Naive \ (green) and ODIWI with $L=10$ iterations (red) estimators.
   The mean (solid line), and 95\% range (ribbon) of 100 replications are presented.
  }
  \label{fig:sim2}
\end{figure}

Figure~\ref{fig:sim2} compares the bias in $\hat \beta_{\hat  x}$ between the ODIWI and \Naive \ estimates, when $\beta_{x}$ change from 0 to 2.
It can be seen that when the true EE is stronger, i.e., the relation between $x$ and $y$ is less linear, there is more to benefit from optimal-design, and so ODIWI's estimates have lower bias than those of the \Naive.

\subsection{Pollution and Glucose Example} \label{sec:empirical}
Recent studies found positive association between PM\textsubscript{2.5} and fasting blood glucose \citep{park2014ambient, peng2016particulate}.
We now employ Algorithm~\ref{algo:iterator}, to re-estimate the EE of PM\textsubscript{2.5} on the potential risk for hyperglycemia.
We are interested in comparing Algorithm ~\ref{algo:iterator}'s estimates to the current (non-iterative) standards in the field.

Exposure data includes daily PM\textsubscript{2.5} measurements over the years 2003-2012 from 46 monitoring stations in Israel, alongside satellite measurements of AOD, normalized-difference-vegetation-index (NDVI), and other spatial and temporal features, which are also available at 1 $km^2$ resolution in the residences of the individuals in the study.

The epidemiological data, also used by \citet{yitshak2016association}, includes over 0.5 million blood glucose tests performed by approximately 43,000 individuals during the years 2003--2012 in Southern Israel, along with other subject's characteristics.
The study was approved by the IRB committee of the Soroka University Medical Center.

We now describe the analysis setup.
We follow \citet{shtein2018estimating} and predict PM\textsubscript{2.5} at the subjects' residences with a linear-mixed-model using satellite measurements and other geospatial features.

The assumed second-stage epidemiological model has the following form:
\begin{align}
\log{\frac{P(y=1|z,\hat x)}{1-P(y=1|z,\hat x)}} = \beta_0 + \sum_{k=1}^6 \beta_{z,k} z_k + \beta_{\hat x} \hat x, 
\end{align}
where $y\in \{0,1\}$ is a binary response indicating whether the subject's blood glucose level is above 126 mg/dl (a clinical cutoff used in diagnosing diabetes); $\beta_0$ is an intercept; 
$z_1,...,z_6$ are covariates capturing seasonal variables and subject's health and socio-economic characteristics (such as age, smoking status, diabetes status, weight, BMI, and socio-economic group), with corresponding coefficients $\beta_{z,1},...,\beta_{z,6}$; 
$\hat x$ is the subject's average predicted exposure to PM\textsubscript{2.5} over last 21 days before test, and $\beta_{\hat x}$ is the corresponding EE.

\begin{figure}[H] 
\centering
  \includegraphics[width=0.85\textwidth]{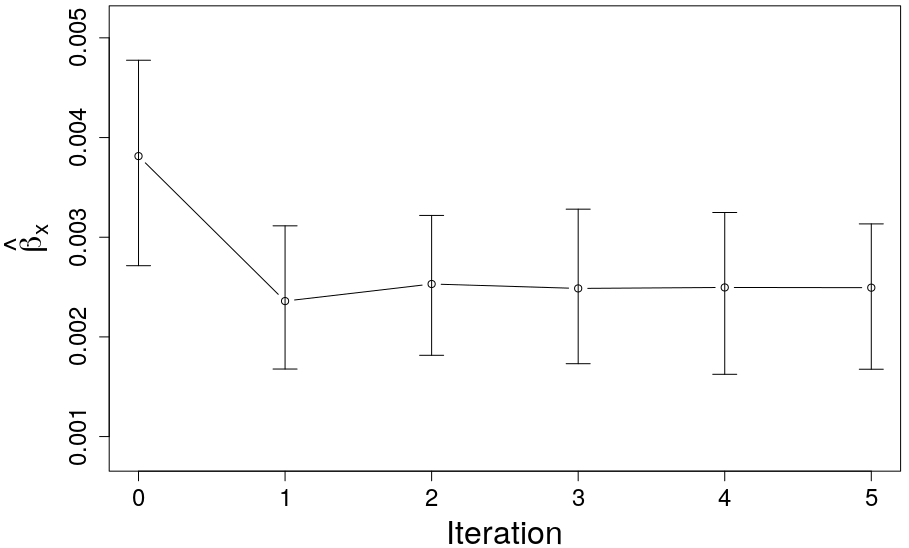}
  \caption{\footnotesize
  PM\textsubscript{2.5} effect on Hyperglycemia risk (log odds ratio): ODIWI estimates and 95\% bootstrap confidence intervals.
  }
  \label{fig:alg_iter_real}
\end{figure}

Figure~\ref{fig:alg_iter_real} presents the progress of Algorithm~\ref{algo:iterator}.
The algorithm was initialized on the \Naive\ estimates (iteration 0), and was operated with $L=5$ iterations.
We report bootstrap confidence intervals, since the usual large-sample---parametric-inference for GLMs does not account for uncertainty introduced by the iterative imputations. 
It can be seen that EE estimates stabilize after two iterations.
Our estimates indicate that the effect of PM on subject's blood glucose level is slightly smaller than the effect estimated in the \naive\ approach, with slightly tighter intervals.

For sensitivity analysis, we repeated this analysis among different subgroups, stratifying along diabetes status or subjects' age.
We also checked the sensitivity of the results to specific confounding variables such as socio-economic or smoking status.
Our findings suggest that a stabilization of the EE estimate after few iterations, on lower values with tighter intervals, is not sensitive to a specific stratification.

\section{Discussion} \label{sec:discussion}
Motivated by the calls of leading epidemiologists and exposure scientists such as \citet{diao2019methods, szpiro2011does}, we tried to improve EE estimates by tailoring predictions to the epidemiological task downstream.
We use optimal-design theory to identify observations of importance, and domain-adaptation theory to improve predictions for those observations. 
Our simulations confirm the validity of this argument: our EE estimates are indeed more accurate than the non-iterative current best practice. 
An empirical demonstration yields smaller EE estimates, with tighter confidence bands. 
This is merely a preliminary demonstration, which is promising: the estimator does make a difference, but it does not completely invalidate previous methods and results. 

Our exposure predictions are less accurate, on average, compared to non-iterative predictions. 
The paradox that worse predictions may improve EE estimates was reported, for instance, in \citet{szpiro2011does}.
Casting the problem using optimal-design theory, perfectly explains this paradox.

\subsection{A Conceptual Difficulty: Exposures Vary with Each New Study}
The reader may wonder whether it is reasonable to let the exposure, a fixed quantity in reality, to vary from (second-stage) study to study.
We argue in favor: allowing the exposure model (and hence the predictions) to vary between studies is not unique to our two-stage setup.
For instance, in supervised-learning, the user is free to make different design decisions that will return different predictions. Namely: choosing a loss function, averaging loss versus median loss, etc. 

Another example is due to ``resolution'': even when global-scale models exist, researchers may prefer country-level predictions.
These, implicitly, prioritize certain areas over others.

We thus argue that using different estimators for different tasks is not unprecedented. 
We merely state explicitly which is the estimation task at hand.

\subsection{A Practical Difficulty: Iterating between Statistician and Epidemiologist}
Non-iterative two-stage studies are the current standard in EEPI. 
Some of the technical and other barriers that made them so popular no longer exist. 
Storage, communication and computing technologies are less stringent.
Unprecedented information sharing is now possible with cloud technology (e.g., ``Copernicus DIAS'' service {\url{https://www.copernicus.eu/en/access-data/dias}}).
These improvements make our iterative approach feasible.

It is also possible to alleviate the computational burden by predicting only exposures that are required for each  study, instead of predicting an entire spatio-temporal domain.
Reducing first-stage's training set may also be justified.

In addition, our estimator does not require the epidemiologist to share data, which may be sensitive, but only the importance preferences. 
This ensures that the two-stage decentralized nature is maintained, and eases privacy concerns.

\subsection{Unprecedented Exposures}
It may be possible that an optimal design will return exposures that have never been seen in the data. 
We thus recommend adding a further restriction: that the optimal design be restricted to the convex-hull of data-points.

\subsection{The number of Support Points}
Choosing the number of support points, $ J $, is a non-trivial matter.
A careful discussion is given by \citet[][Sec. 8.3]{pukelsheim2006optimal}.
We rely on \citet{fedorov2013optimal} who state a simple and usually satisfied condition under which the maximum number reduces to $|\Phi|(|\Phi|+1)/2$.

\subsection{The Effect of Second-Stage Covariates} \label{sec:second-stage-covariates}
First-stage data does not include personal covariates that are available at the second stage. 
This means that the optimal design may not depend on such covariates. 
We thus suggest either ignoring covariates when finding an optimal design, or optimizing predictions for a particular value of the covariates such as their median.

\subsection{Generalization form Pollution to other Exposures}
In this presentation we focused on satellite measurements of air-pollution and temperature, but the ideas are not limited to those examples.
Versions of Algorithm~\ref{algo:iterator} may be applied in any other two-stage study.
Two-stage studies are used in environmental epidemiology for other exposures such as NOx, O3, air pollen, light at night, etc.
Two-stage studies are also used in other epidemiological sub-fields such as nutritional epidemiology, occupational epidemiology, and more \citep{szpiro2013measurement, wu2019methods}.
All these fields may gain accuracy by using optimal-design and domain-adaptation theory to guide predictions where they truly matter.

\subsection{Departing from GLMs} \label{sec:departing-from-glm}
When the second stage is a linear model, our approach can be simplified.
Unlike GLMs, in linear models information is maximized when sampling at the boundary of the design space, and independently from the unknown effects.
This suggests that for second-stage linear models, it may be enough to provide good predictions for extreme exposure values, without iterating. 
Other than GLM, optimal-design theory exists for many other nonlinear second-stage models, including survival and longitudinal models, \citep{dean2015handbook, fedorov2013optimal}.

\subsection{When to Use the Estimator?}
Two major components are required for our estimator: 
(1) Second-stage optimal-design theory. 
(2) First-stage domain-adaptation. 
The stronger the second-stage non-linearities, then more there is to gain from our iterations. 

Caution should be taken if the second-stage model is misspecified, as estimated EE affects predicted exposures and vice-versa.
Bias in the second stage may introduce bias in the first stage, a phenomenon known in the EEPI literature as \emph{feedback} \citep{sheppard2012confounding}.

\subsection{User Selected Tuning Parameters} \label{sec:disc_algo}
In this section we discuss some design choices that can be made in Algorithm~\ref{algo:iterator}.

\subsubsection{Initialization}
A natural importance weights initiation is uniform weights: $\omega^0(\x_i) := 1/ \n$.
We did however find that averaging $\hat \beta$ over multiple random initializations of $\omega^0$ is beneficial.
This is because the two-stage estimator has many degrees of freedom, and is prone to overfitting. 
To see why this is the case, consider a null effect, $ \beta_x=0 $.
Because our estimator sequentially updates the design, an error in the initial estimate, $ \hat \beta^1 $, will affect downstream iterations (feedback effect).
We found that aggregating multiple initializations alleviate this feedback. 

There are many ways to aggregate the multiple $\hat \beta$. 
We considered two: average estimates after a single iteration (then continue serially), or average after the last iteration. 
Our simulation results suggest there is no significant difference in the estimation accuracy between the two.

\subsubsection{Over-fitting}
Too many iterations can lead to overfitting of the training data.
Unlike usual supervised learning problems, where one can hold a validation set to alarm when performance stops improving, in estimation problems, a holdout dataset does not protect from bias.

Yet, there are some choices in the optimization process we can make that reduce overfitting.
One such is defining a low learning rate between iterations.
A momentum method for $\hat \beta$ is suggested: $\check \beta^l = \alpha \check \beta^{l-1} + (1-\alpha) \hat \beta^{l}$.
Choosing higher $\alpha$ dampens oscillations in $\check \beta$, and is more safe.

The kernel operator, $ \K $, in Eq.~\ref{eq:PT} is another design choice that governs the tendency to overfit.
Our simulation results suggest that the choice of the kernel function is less significant (we compared uniform, Gaussian, and triangle).
The kernel's bandwidth, on the other hand, is more influential. 
A wider bandwidth means slower convergence, and is more safe in general.

When the choice of number of iterations may be critical, in the spirit of \emph{one-step estimators} \citep{bickel1975one}, we advocate a single iteration from each initialization ($ L=1 $). 
Setting $ L=1 $ may be suboptimal if $ \m $ is highly non-linear.
The fact that accuracy is improved after a single iteration, is confirmed in our simulation analysis (Section~{\ref{sec:sim}}).

\subsubsection{Optimal-Design Criterion}
In Algorithm~\ref{algo:iterator} we used local D-optimality because of its popularity and computational convenience.
Our simulation results suggest the estimates are insensitive to the optimality criterion such as E-optimality, A-optimality, etc, but with multivariate exposures this may change.

\subsection{Prior Shift Assumption} \label{sec:prior-shift}
The prior-shift assumption, $P_S(r|x) = P_T(r|x)$, means that the distribution of geographical covariates given an exposure, does not vary between first and second stage.
Put differently, the difference between the joints is only caused by a change in the exposure marginal distribution.
This assumption will often not hold, but Algorithm~\ref{algo:iterator} may still remain useful.

We used the prior-shift assumption to derive the weights in Eq.\ref{eq:iw}: putting more importance on samples with similar exposures as those of the important samples in the second stage.
This was merely a construction device.
Our simulation shows that estimates may improve even when this assumption is invalidated.
Moreover, a domain-adaptation of the first to the second stage can be achieved without this assumption using other methods \citep{weiss2016survey}.
For instance by estimating $P(r,x)$ in both stages directly, using prior knowledge on the important individuals (e.g., about their residence).

\subsection{Future Research}
This work is merely a proof of concept in an attempt to answer the call of \citet{diao2019methods} and \citet{szpiro2011does}.
A lot of work is still required on the statistical properties of the proposed estimator, recommended usage, sensitivity analysis of its assumptions, further comparisons with the non-iterative approach, and also with the full two-stage likelihood approach.

\bibliographystyle{apalike}
\bibliography{refs.bib}

\newpage

\end{document}